# Commensurate and Incommensurate Chern Insulators in Magic-angle Bilayer Graphene

Zaizhe Zhang[1†], Jingxin Yang[1,2†], Bo Xie[3†], Zuo Feng[4], Shu Zhang[5], Kenji Watanabe[6], Takashi Taniguchi[7], Xiaoxia Yang[5], Qing Dai[5,8], Tao Liu[2], Donghua Liu[9], Kaihui Liu[4], Zhida Song[1,10,11], Jianpeng Liu[3] and Xiaobo Lu[1,10*]

[1]International Center for Quantum Materials, School of Physics, Peking University, Beijing 100871, China
[2]National Engineering Research Center of Electromagnetic Radiation Control Materials, University of Electronic Science and Technology of China, Chengdu 611731, China
[3]School of Physical Science and Technology, ShanghaiTech University, Shanghai 201210, China
[4]State Key Laboratory for Mesoscopic Physics, Frontiers Science Centre for Nano-optoelectronics, School of Physics, Peking University, Beijing 100871, China
[5]CAS Key Laboratory of Nanophotonic Materials and Devices, CAS Key Laboratory of Standardization and Measurement for Nanotechnology, CAS Center for Excellence in Nanoscience, National Center for Nanoscience and Technology, Beijing 100871, China
[6]Research Center for Electronic and Optical Materials, National Institute of Material Sciences, 1-1 Namiki, Tsukuba 305-0044, Japan
[7]Research Center for Materials Nanoarchitectonics, National Institute of Material Sciences, 1-1 Namiki, Tsukuba 305-0044, Japan
[8]School of Materials Science and Engineering, Shanghai Jiao Tong University, Shanghai, 200240, China
[9]School of Materials and Energy, University of Electronic Science and Technology of China, Chengdu 611731, China
[10]Collaborative Innovation Center of Quantum Matter, Beijing 100871, China
[11]Hefei National Laboratory, Hefei 230088, China

* E-mail: xiaobolu@pku.edu.cn
† These authors contributed equally to this work.

**The interplay between strong electron-electron interaction and symmetry breaking can have profound influence on the topological properties of materials[1–4]. In magic angle twisted bilayer graphene (MATBG), the flat band with a single SU(4) flavor associated with the spin and valley degrees of freedom gains non-zero Chern number when $C_{2z}$ symmetry or $C_{2z}T$ symmetry is broken. Electron-electron interaction can further lift the SU(4) degeneracy, leading to the Chern insulator states[5–14]. Here we report a complete sequence of zero-field Chern insulators at all odd integer fillings ($v = \pm1, \pm3$) with different chirality ($C = 1$ or -1) in hBN aligned MATBG which structurally breaks $C_{2z}$ symmetry. The Chern states at hole fillings ($v = -1, -3$), which are firstly observed in this work, host an opposite chirality compared with the electron filling scenario. By slightly doping the $v = \pm3$ states, we have observed new correlated insulating states at incommensurate moiré fillings which is highly suggested to be intrinsic Wigner crystals according to our theoretical calculations. Remarkably, we have observed prominent Streda-formula violation around $v = -3$ state. By doping the Chern gap at $v = -3$ with notable number of electrons at finite magnetic field, the Hall resistance $R_{yx}$ robustly quantizes to ~ $h/e^2$ whereas longitudinal resistance $R_{xx}$ vanishes, indicating that the chemical potential is pinned within a Chern gap, forming an incommensurate Chern insulator. By providing the first experimental observation of zero-**

field Chern insulators in the flat valence band, our work fills up the overall topological framework of MATBG with broken $C_{2z}$ symmetry. Our findings also demonstrate that doped topological flat band is an ideal platform to investigate exotic incommensurate correlated topological states.

Flat band in magic-angle twisted bilayer graphene (MATBG) offers an ideal platform to study novel band topology as well as other quantum phases including superconductivity, correlated insulators and orbital magnetism[6–13,15–32]. From single-particle perspective, the lowest flat valence and conduction bands of MATBG touch at Dirac points, which are protected by $C_{2z}T$ symmetry. Breaking either $C_{2z}$ symmetry or $C_{2z}T$ symmetry can gap out the Dirac point and lead to a net Chern number of $C=1$ or $C=-1$ for each flat band with a single SU(4) isospin flavor[5,13]. Past studies[8–10] have shown that the $C_{2z}$ symmetry of graphene can be effectively broken by making lattice alignment with hexagonal boron nitride (hBN) and Chern insulator at zero magnetic field have been observed with moiré filling $v = +3$ (three electrons filled in moiré unit cell). Later studies[11,12] have shown that the strong electron correlation in MATBG could also break the $C_{2z}T$ symmetry even without alignment to hBN substrates, giving rise to different Chern insulating states, including $v = +1$ (one electrons filled in moiré unit cell) state with a Chern number of $C = -1$ robust down to zero magnetic field as inferred from Streda formula behavior of $R_{xx}$ minima, and a sequence of Chern insulators onset at finite magnetic fields.

However, so far Chern insulating states at zero magnetic field characterized by (quantized) anomalous Hall effect in MATBG have only been observed in the conduction band, and the quantized $R_{xy}$ has been only observed at filling $v = +3$. The topological nature of valence flat band is still not well understood. Theoretically, the exact topological properties of the correlated states in MATBG highly rely on the interlayer hopping parameters, i.e., $w_0$ for AA interlayer hopping and $w_1$ for AB interlayer hopping[5], which are also closely related to other quantum phases including superconductivity and magnetism in this system[33]. Experimentally mapping out the overall topological framework in particular within a single MATBG device will be inspirational, but also challenging due to the difficulty for controlling twist angles and the fragility of these Chern insulating states.

In this work, we report the observation of multiple commensurate Chern insulator states with opposite chiralities on the electron and hole sides in a same MATBG device. Most saliently, incommensurate Chern insulators which clearly violate Streda formula behavior (as expected from conventional Landau levels) have been observed in MATBG aligned to hBN substrate (Fig. 1a).

**Complete sequence of Chern insulators**

As shown in Fig. 1b, the staggered sublattice potential induced by the aligned hBN substrate can effectively break the $C_{2z}$ symmetry, leading to the emergence of opposite valley Chern numbers for the flat bands. At zero magnetic field, the Chern insulator states can be spontaneously polarized into either valley driven by interactions, since the flat bands of the two opposite valleys are energetically degenerate at the non-interacting level. Given the same valley polarization, Fig. 1b shows the filling configurations of the flat band with single isospin flavor for different Chern states observed in our experiment. Fig. 1c shows $R_{xx}$ of device D1 (twist angle $\theta \sim 1.15°$) as a function of temperature and moiré filling factor $v$. Similar with previously reported results, resistive correlated states at integer moiré fillings can be resolved[15–18]. As a result of substrate induced $C_{2z}$ symmetry breaking, the charge neutrality point (CNP) exhibits an unprecedentedly large gap ($\Delta_{CNP} \approx 21.15$ meV) with clear thermal activation shown in the inset of Fig. 1c. Fig. 1d shows anti-symmetrized Hall resistance $R_{yx}$ as a function of filling factor $v$ (fast axis and sweeping from left to right) and perpendicular magnetic field $B_\perp$ measured at a based phonon temperature of $T \sim 10$ mK. When the carrier density is close to values associated with odd integer fillings ($v = \pm 1, \pm 3$), $R_{yx}$ undergoes an abrupt sign reversal at zero magnetic field which is further illustrated in Extended Data Fig. 1 and highly implying the formation of zero-field Chern insulators. In conduction band, the $R_{yx}$ values at $v = 1$ and $v = 3$ can reach ~ 70% and 95% of one quantized resistance value ($h/e^2$) at $B_\perp = 480$ mT (Extended Data Fig. 1e), respectively, which is consistent with previous work. Interestingly, we firstly observe zero-field Chern insulators at the flat valence band of MATBG with hall resistance reaching 98% for $v = -3$ state and 89% for $v = -1$ state at $B_\perp = 480$ mT. As clearly shown in Fig. 1d, Chern insulator for $v = -3$ ($v = -1$) shows opposite Chern number with the $v = 3$ ($v = 1$) state (when $B_\perp > 300$ mT). These experimental results are in good agreement with our Hartree-Fock calculations (Method and Extended data Fig. 2), which further confirm the exact topological ground states in MATBG with $C_{2z}$ symmetry breaking. To shortly summarize, the magnetization of MATBG which is dominated by orbital magnetism is energetically favored to be align with the external $B_\perp$ field. At finite $B_\perp$ field, as shown in Fig. 1b, the Chern insulator states at valence band and conduction band are polarized to the same valley (same magnetization) but with opposite Chern numbers.

It is noteworthy that in Fig. 1c there is additional sign reversal of $R_{yx}$ for $v = 3$ state at $B_\perp \approx \pm 300$ mT. Careful measurements show that the sign change of $R_{yx}$ sensitively depends on the sweeping direction of $v$. As shown in Extended data Fig. 3 which is measured with opposite sweeping direction of $v$ (from right to left), the sign reversal behavior of $R_{yx}$ appears at $v = -3$. Similar sign reversal of $R_{yx}$ has also been reported at $v = 1$ state and other systems showing Chern insulators[11,34–38] and can be attributed to the competition between the two component of orbital magnetization including Chern magnetization contributed by the gapless edge states and magnetization from self-rotation of the wavepacket[11,39]. We also note that our device does not exhibit zero-field Chern insulators at $v = \pm 2$, indicating the absence of valley polarization for $v = \pm 2$ states[36]. Furthermore, topologically trivial charge density wave (CDW) states can be resolved at fractional moiré fillings (Extended Data Fig. 1c and d). The $v = -2/3$ state starts to appear at $B_\perp = \pm 0.5$ T whereas $v = 5/3$ state can be stabilized at $B_\perp = \pm 0.2$ T.

To further investigate the property of Chern insulators, we measured the magnetic response for these states with fixed carrier density. In Fig. 2a-d, we display the hysteresis loops of $R_{yx}$ and $R_{xx}$ as the magnetic field is swept back and forth at different moiré fillings, showing Chern insulators with $R_{yx}$ values at zero magnetic field reaching 90.2% ($v = -3$), 83.5% ($v = -1$), 63.7% ($v = 3$), and 91.3% ($v = 3$). The imperfect quantization of $R_{yx}$ and non-zero $R_{xx}$ is due to non-uniform magnetic domains and concomitant percolation in Chern insulator[11]. We note that the additional sign change of $v = \pm 3$ states is missing in the magnetic hysteresis loop wherein the chemical potential and magnitude of the magnetization are fixed. Notably, the firstly observed Chern insulators at $v = -3$ and $v = -1$ states exhibit magnetic hysteresis loops with chirality exactly opposite to those in $v = 3$ and $v = 1$ states, which is consistent with our Hartree-Fock calculation showing opposite Chern numbers between the Chern insulators with $v > 0$ and $v < 0$, given the same valley polarization

(Extended Data Fig. 2). And most of these features are reproduced in different contact pairs of device D1 (Extended Data Fig. 4) and device D2 (Extended Data Fig. 5).

Fig. 2e-h illustrate the temperature dependence of magnetic hysteresis loops measured at $v = \pm 1$ and $v = \pm 3$ states in device D1. Both $v = \pm 3$ states exhibit comparable Curie temperatures which are defined by the point where the magnetic hysteresis loops of $R_{yx}$ vanish. As shown in Fig. 2e and h, the Curie temperatures are around $T_c = 2.0$ K for $v = -3$ and $T_c = 2.4$ K for $v = 3$. Interestingly, the $v = +1$ state with $T_c = 10.7$ K and $v = -1$ state with $T_c = 8.6$ K have notably higher Curie temperatures than $v = \pm 3$ states. Whereas, $v = \pm 1$ states exhibit less perfect quantization of $R_{yx}$. Similar behaviors have also been observed in the measurement of $R_{xx}$ as a function of temperature $T$ and filling factor $v$ under zero $B_\perp$ field (Extended data Fig. 6). The abrupt change of $R_{xx}$ with decreasing temperature indicates the emergence of orbital magnetism along with the formation of Chern gaps.

**Incommensurate correlated insulators**

Fig. 3a and b show the longitudinal resistance $R_{xx}$ as a function of temperature $T$ and moiré filling factor $v$ close to the $v = \pm 3$ states. Below the Curie temperature ($T_c \approx 2.0$ K), Chern insulator accompanied by orbital magnetism emerges which are evidenced by the vanishing $R_{xx}$ value and nearly quantized hysteretic $R_{yx}$ shown in Fig. 2. Interestingly, unexpected phenomena have been observed at slightly electron doped $v = -3$ state and hole doped $v = 3$ state, showing a highly resistive dome at $v = -3+\delta$ and $v = 3-\delta$. Fig. 3e displays the temperature-dependent resistance $R_{xx}$ near $v = -3$ for a series of moiré fillings, exhibiting notable transition from the Chern insulator with metallic $R_{xx}$ to incommensurate states with insulating $R_{xx}$. Similar behaviors near $v = 3$ state have been observed and shown in Extended Fig. 7. Detailed measurements have further revealed insulating behavior of the resistive dome with the maximum gap size $\Delta \sim 0.12$ meV at $v = 2.89$ (Fig. 3d). Fig. 3c schematically illustrates different correlated phases near $v = -3$ extracted from Fig. 3a, showing a resistive correlated state (CS) at $v = -3$ above Curie temperature, a Chern insulator (ChI) at $v = -3$ below Curie temperature and an incommensurate correlated insulator (ICI) dome at $v = -3+\delta$. The inset of Fig. 3c shows the calculated filling-dependent Wigner-Seitz radius $r_s$ which characterizes the ratio of Coulomb repulsion to kinetic energy[40]. The $r_s$ value increases rapidly close to $v = -3$ and $v = 3$ (Extended Fig. 7c), reaching $r_s > 40$ at the moiré fillings showing ICI. In two-dimensional electron systems, different $r_s$ values can lead to different ground states. For instance, quantum Monte Carlo calculation indicates that the formation of Wigner crystals needs $r_s > 31$[41,42]. However, it is challenging to achieve such a high value of $r_s$ without magnetic field which depends on different parameters, $r_s = m^*e^2/4\pi\hbar^2\varepsilon_0\varepsilon_r(\pi n)^{1/2}$, where $m^*$, e, $n$, $h$, $\varepsilon_0$ and $\varepsilon_r$ denote the effective electron mass, elementary charge, carrier density, reduced Planck constant, vacuum and relative permittivity. For instance, considering a two-dimensional electron system with $\varepsilon_r = 13$ and $m^* = 0.07m_0$ where $m_0$ is the free electron mass (parameters taken from GaAs two-dimensional electron gas, 2DEG), $r_s > 31$ requires a carrier density $n < 3.4 \times 10^8$ cm$^{-2}$ which is extremely challenging to be resolved experimentally. The appearance of ICI states at $v = -3+\delta$ and $v = 3-\delta$ with ultra-high $r_s$ values is mainly due to the large effective mass $m^*$, which is also consistent with the heavy fermion picture developed recently[43–49]. In sharp contrast with the generalized Wigner crystals widely reported in a fractionally filled moiré band[50–54], the Wigner-crystal-like ICI states spontaneously break the translational symmetry and provide a new type of insulators in MATBG beyond those lifting the SU(4) flavor degeneracy.

**Incommensurate Chern insulators**

Exotic phenomena have been observed in the $B_\perp$ dependent behavior of the ChI at $v = -3$. Generally, the Hall resistance $R_{yx}$ changes its sign at finite $B_\perp$ field when the chemical potential is tuned across a topologically trivial gap. For a Chern gap scenario, the evolution of the gap in $n$-$B_\perp$ ($n$ is the carrier density) parameter space is further expected to follow Streda formula $\frac{dn}{dB_\perp} = \frac{Ce}{h}$ where $C$ is the Chern number, $e$ is the electron charge and $h$ is Planck's constant. Strikingly, the behavior of $v = -3$ state phenomenally violates Streda formula, with neither a sign reversal of $R_{yx}$ at $v = -3$ nor a standard trajectory following the slope of $\frac{dn}{dB_\perp} = -\frac{e}{h}$ has been observed. Instead, the $v = -3$ state persists over a very board region (rendered with dark blue in Fig. 4a) with $C = -1$ in $v$-$B_\perp$ parameter space (Fig. 4a and b). Fig. 4d further quantitively shows the quantized $R_{yx}$ and vanishing $R_{xx}$ at $B_\perp = 5$ T for the Chern gap from stemming $v = -3$ and the conventional Landau level (LL) gaps from $v = -2$. In Fig. 4c, the schematic shows the broad $C = -1$ state at $v = -3$ as well as the different LL gaps fanning out from $v = -2$. For the $C = -1$ state at $v = -3$, the red solid line qualitatively indicates an imaginary boundary between the normal Chern gap (leftward) which follows the Streda formula and electron doped Chern gap (rightward) which robustly shows quantized $R_{yx}$ but violates Streda formula. We note that the significant broadening feature only exists near $v = -3$ state whereas other Landau level gaps strictly follow the Streda formula, which can exclude the dominant origin of disorder effect. Furthermore, the two cornners indicated by red circles in Fig. 4a can be accurately traced to $v = -8/3$ and $-7/3$ at zero $B_\perp$ field with a trajectory corresponding to $C = -1$, indicating the formation of topological states with indices $(v, C) = (-8/3, -1)$ and $(-7/3, -1)$ which are commensurate to moiré potential. The absence of phase boundary between the commensurate states and incommensurate states demonstrates that the additionally doped electrons spontaneously break the translational symmetry and condense into localized bulk states. As shown in Fig. 4d, the localized states start to contribute conductivity with more electrons doped (pink region). One possible reason is that the localized states start to melt with more electrons doped. In addition, similar states with prominent Streda-formula violation have been observed proximate to the (-8/3, -1) state, showing $C = -2, -3$, (Extended Fig. 8) etc. These states stabilized at finite field can be attributed to the Landau level excitations from $v = -8/3$.

The topologically trivial properties of $v = -2$ state indicate that the state is valley unpolarized. Two flat valence bands from different valleys with opposite valley Chern numbers are filled at $v = -2$ state, leading to a net Chern number of zero. For the ChI state at $v = -3$, only a single-flavor flat band is filled as shown in the bottom panel of Fig. 5a with quantitative band structure shown in Fig. 5e (More details shown in Method). Fig. 5a qualitatively shows different microscopic pictures for different quantum phases shown in Fig. 5b. When the $v = -3$ state is slightly doped with electrons at zero magnetic field, the flat band at $K'$ valley starts to be filled (middle panel of Fig. 5a). These electrons will be localized in the AA regions due to the large $r_s$ values originated from large effective mass [43-49] and low carrier density (Fig. 3c). Fig. 5c and d show the calculated distribution of electron pockets of the slightly filled $K'$ valley at $v = -2.9$ on top of energy dispersion and Berry curvature distribution in the momentum space, respectively. Interestingly, these electron pockets have vanishing Berry curvature, as Berry curvature is mainly concentrated at $\Gamma_s$ point (Fig. 5d). This gives rise to the zero-field IChI state which maintains the topological property of $v = -3$ state. With more electron doped, the zero-field IChI state will be replaced by a valley unpolarized

ICI state with large $r_s$ values and zero net Chern number (upper panel of Fig. 5a). As $B_\perp$ field energetically favors both spin and valley polarization, the valley unpolarized ICI state will transform into IChI state with an external $B_\perp$ field, leading to a fully filled flavor at one valley and slightly filled flavor at the other valley (upper panel of Fig. 5a). Given the robust extended $C = -1$ state and the large $r_s$ values near $v = -3$, one of the possible ground state is Wigner crystal combined with a Chern insulator, which is reminiscent of WCs around integer Landau fillings in 2DEG[55]. Fig. 5f schematically shows the possible real-space picture of IChI state at $v = -3+\delta$ and ChI state at $v = -3$. For the ChI state, each moiré unit cell is filled with one electron indicated by the red dots in Fig. 5f, forming a commensurate correlated insulators and non-dissipative edge state. For the IChI states, the additionally doped electrons in the other valley could condense into Wigner crystals (yellow dot in Fig. 5f) which are further pinned by moiré potential or disorders, whereas the topological properties inherit from $v = -3$ state with $C = -1$.

**Discussion**

Our work for the first time demonstrates the full sequence of zero-field Chern insulators which are theoretically predicted in MATBG aligned to hBN. At incommensurate fillings, we have observed both topologically trivial and nontrivial correlated states (ICI and IChI states) extended over a broad region in the $v$-$B_\perp$ space, by virtue of ultra large $r_s$ values. We speculate the partially valley polarized IChI can also exist in other systems, i.e., rhombohedral multilayer graphene moiré superlattices[56,57], when the $r_s$ values are large enough. Additional insight into the IChI state can be obtained from frequency-dependent transport measurement, which can reveal its dynamic behavior. Another promising extension of our work would be scanning probe microscope measurement of the system to show the smoking-gun evidence for the possible Wigner crystallization.

Acknowledgements:
We are grateful for fruitful discussions with B. Andrei Bernevig, Dmitri K. Efetov, Biao Lian, Xi Dai, Yang Liu and X.C. Xie. X.L. acknowledges support from the National Key R&D Program (Grant Nos. 2022YFA1403502) and the National Natural Science Foundation of China (Grant Nos. 12274006 and 12141401). K.W. and T.T. acknowledge support from the JSPS KAKENHI (Grant nos. 21H05233 and 23H02052) and World Premier International Research Center Initiative (WPI), MEXT, Japan. T.L. acknowledges support from the National Natural Science Foundation of China (Grant Nos. 52072060).

Author contributions:
X.L. and Z.Z. conceived and designed the experiments; Z.Z. and J.Y. fabricated the devices and performed the measurement with help from Z.F. and S.Z.; Z.Z., Z.S., J.L. and X.L. analyzed the data; J.L. and B.X. performed the theoretical modeling; T.T. and K.W. contributed materials; X.L., Q.D., X.Y., K.L., T.L., and D.L. supported the experiments; Z.Z., B.X., J.L. and X.L. wrote the paper.

## Methods

**Device fabrication.**

Our devices are fabricated using a standard 'cut-and-stack' dry transfer method. Graphene, graphite and hBN are exfoliated on $O_2$ plasma cleaned (30 W, 2 mins) $Si^{++}/SiO_2$ (285 nm) chips. The graphene flake is cut into several pieces by a femtosecond laser (central wavelength ~ 517 nm, pulse width ~ 150 fs, repetition frequency ~ 80 MHz and maximum power ~ 150 mW). A PC (poly bisphenol A carbonate) / PDMS (polydimethylsiloxane) stamp was used to pick up top hBN, two pieces of graphene, bottom hBN and back gate sequentially, the crystalline edges of the graphene and hBN flakes are meticulously aligned during the transfer process, the whole heterostructure was then released onto a $Si^{++}/SiO_2$ substrate when the PC was backed to 180 °C. Finally, the device was defined into the Hall bar geometry utilizing standard electron beam lithography (EBL) and reactive ion etching (RIE) techniques, and the edge contacts (5 nm Cr / 70 nm Au) were formed using a combination of electron beam evaporation and thermal evaporation techniques.

**Electrical transport measurement.**

Electrical transport measurements were performed within a dilution refrigerator, maintaining a base phonon temperature of approximately 10 mK. Keithley 2400 source-meters were used to apply bottom gate voltage. Stanford Research Systems SR860 lock-in amplifiers were employed to measure the four-terminal longitudinal resistance ($R_{xx}$) and Hall resistance ($R_{yx}$) using an AC current bias ranging from 0.1 to 10 nA at a frequency of 7.777 Hz, and the $R_{xx}$ and $R_{yx}$ were amplified utilizing a Stanford Research Systems SR560 voltage preamplifier.

**Twist angle extraction.**

To determine the twist angle, we use the relationship $n_s = 8\theta^2/3^{1/2}a^2$ at small twisted angle limit, where $\theta$ represents the interlayer twist angle of twisted bilayer graphene, $n_s$ denotes the charge carrier density corresponding to the fully filled a superlattice moiré unit cell, the value of $n_s$ is inferred from the $n_s = C_b V_s/e$, where $V_s$ represents the voltage value of the back gate corresponding to full filling peaks, $C_b$ denotes the back gate capacitance, the values of $C_b$ are determined by the Landau fan diagram, and $a = 0.246$ nm signifies the interatomic distance in monolayer graphene. Then we convert the charge carrier density $n$ to the moiré filling factor $v$ ($v = 4n/n_s$) by utilizing a series of correlated insulating states corresponding to distinct longitudinal resistance $R_{xx}$ peaks, and we derive the twist angle in our device D1 as $\theta = 1.15°$ and $\theta = 1.09°$ for device D2.

**Theoretical calculation model.**

We investigate the electronic properties of magic-angle twisted bilayer graphene (TBG) at integer fillings of the flat bands. We start with the famous Bistritzer-MacDonald continuum model[32] to depict the low energy effective Hamiltonian for TBG in the valley $\mu$ (for $K/K'$ valley):

$$H_\mu^0 = \begin{bmatrix} -\hbar v_F(\mathbf{k} - \mathbf{K}_1^\mu) \cdot \sigma_\mu + V_{hBN} & U_\mu(\mathbf{r}) \\ U_\mu^\dagger(\mathbf{r}) & -\hbar v_F(\mathbf{k} - \mathbf{K}_2^\mu) \cdot \sigma_\mu \end{bmatrix},$$

where $v_F$ denotes the Fermi velocity, $\mathbf{K}_{1,2}^\mu$ represent two Dirac points from $\mu$ valley, and $\sigma_\mu = (\mu\sigma_x, \sigma_y)(\mu = \pm)$ are Pauli matrices in the sublattice space. The $U_\mu(\mathbf{r})$ matrix describe the moiré potential between two graphene layers. The single aligned h-BN substrate introduces an effective moire potential term on the top layer of TBG[58]. In order to simplify our analysis, we assume that the moiré potential induced by the h-BN and the moiré potential induced by the bilayer graphene

share the same periodicity in real space. As a result, the moiré potential induce by h-BN is given by:

$$V_{hBN} = V_0 \begin{bmatrix} 1 & 0 \\ 0 & 1 \end{bmatrix} + \left\{ V_1 e^{i\mu\psi} \left[ \begin{bmatrix} 1 & \omega^{-\mu} \\ 1 & \omega^{-\mu} \end{bmatrix} e^{i\mu G_1^M \cdot r} + \begin{bmatrix} 1 & \omega^{\mu} \\ \omega^{\mu} & \omega^{-\mu} \end{bmatrix} e^{i\mu G_2^M \cdot r} + \begin{bmatrix} 1 & 1 \\ \omega^{-\mu} & \omega^{-\mu} \end{bmatrix} \right] e^{-i\mu(G_1^M + G_2^M) \cdot r} + h.c. \right\},$$

where $\omega = e^{i2\pi/3}$, $V_0 \approx 0.0289$ eV, $V_1 \approx 0.0210$ eV and $\psi \approx -0.29$ (rad).

Then we consider the dominant intravalley component of the long-range Coulomb interaction in TBG system. The interaction Hamiltonian is given by:

$$H_C = \frac{1}{2N_s} \sum_{\lambda\lambda'} \sum_{\mathbf{kk'q}} V(\mathbf{q}) \hat{c}^\dagger_{\mathbf{k+q},\lambda} \hat{c}^\dagger_{\mathbf{k'-q},\lambda'} \hat{c}_{\mathbf{k'},\lambda'} \hat{c}_{\mathbf{k},\lambda},$$

where $N_s$ is the number of moiré supercell in the system, and $\mathbf{k}$ and $\mathbf{q}$ are the wave vector relative to the Dirac points. $\lambda \equiv (\mu, \alpha, \sigma)$ represents the flavor, encompassing the valley ($\mu$), layer-sublattice ($\alpha$) points and spin ($\sigma$) subspace. The Coulomb interaction can be described by: $V(\mathbf{q}) = e^2 \tanh(|\mathbf{q}|d_s)/(2\Omega_M \epsilon_{BN}\epsilon_0|\mathbf{q}|)$. $\Omega_M$ is the area of the moiré supercell, $d_s = 40$ nm, $\epsilon_{BN} = 4$ is the dielectric constant of h-BN, and $\epsilon_0$ is the vacuum permittivity. We employ the Hartree-Fock approximation to the Coulomb interaction and self-consistently solve the Hamiltonian[59] $H_0 + H_C$. We project the interaction Hamiltonian (with Hartree-Fock approximation) onto the flat band subspace. The remote bands below the charge neutrality point are all occupied, which can interact with the electrons in the flat bands. This effect can be described as a remote band potential acting on the flat band subspace[49,60], enhancing the band width of the flat bands to about 50 meV and breaking the particle-hole symmetry. Furthermore, in addition to the screening effect from the metallic gate, the Coulomb interaction can be further screened by the virtual particle-hole excitation from the remote bands. We follow previous researches[59] and characterize such screening effects using the constrained random phase approximation (cRPA).

We perform the unrestricted Hartree-Fock calculations within the low-energy flat band subspace including the remote band potential, with cRPA screened Coulomb interactions. The dominant order parameters in all the cases are $\tau_z$, $s_z$, and $\tau_z s_z$, where $\tau$ and $s$ are Pauli matrix defined in the valley and spin subspace, respectively.

Furthermore, we calculate the Wigner-Seitz radius $r_s$ to characterize the formation of a Wigner crystal state. The Wigner-Seitz radius is defined as: $r_s = m_e^* r_e/(2\epsilon m_0 a_B)$. $r_e$ is the average distance between the electrons in the moiré flat bands, which satisfies $\pi r_e^2 = 1/n_e$, and $n_e$ is the carrier density. $m_e^*$ is the effective mass of electron and $m_0$ is the bare electron mass. $a_B$ is the Bohr radius.

**Symmetrize $R_{xx}$ and anti-symmetrize $R_{yx}$.**

Due to imperfect Hall geometry definition in our devices, there is a mixing of the $R_{xx}$ and $R_{yx}$, and we employed the standard procedure of symmetrizing and anti-symmetrizing the raw measured data to correct the mixing effect. This procedure allows us to obtain accurate values of $R_{xx}^S$ and $R_{yx}^{AS}$ respectively.

When performing measurements to obtain the Landau fan diagrams (fast sweeping axis: $\nu$, slow sweeping axis: $B$), we employ the following symmetrize and anti-symmetrize procedure:

$$R_{xx}^S(B_0, \nu) = \frac{R_{xx,\text{raw}}(B_0, \nu) + R_{xx,\text{raw}}(-B_0, \nu)}{2}, R_{xx}^S(-B_0, \nu) = R_{xx}^S(B_0, \nu)$$

$$R_{yx}^{AS}(B_0, \nu) = \frac{R_{yx,\text{raw}}(B_0, \nu) - R_{yx,\text{raw}}(-B_0, \nu)}{2}, R_{yx}^{AS}(-B_0, \nu) = -R_{yx}^{AS}(B_0, \nu)$$

When performing measurements to obtain the hysteresis loops (with the sweeping axis as $B$, and $v$ fixed at $v_0$), we employ the following symmetrize and anti-symmetrize procedure:

$$R_{xx}^{\uparrow}(B, v_0) = \frac{R_{xx,\text{raw}}^{\uparrow}(B, v_0) + R_{xx,\text{raw}}^{\downarrow}(-B, v_0)}{2}$$

$$R_{xx}^{\downarrow}(B, v_0) = \frac{R_{xx,\text{raw}}^{\downarrow}(B, v_0) + R_{xx,\text{raw}}^{\uparrow}(-B, v_0)}{2}$$

$$R_{xx}^{\uparrow}(-B, v_0) = R_{xx}^{\downarrow}(B, v_0), R_{xx}^{\downarrow}(-B, v_0) = R_{xx}^{\uparrow}(B, v_0)$$

$$R_{yx}^{\uparrow}(B, v_0) = \frac{R_{yx,\text{raw}}^{\uparrow}(B, v_0) - R_{yx,\text{raw}}^{\downarrow}(-B, v_0)}{2}$$

$$R_{yx}^{\downarrow}(B, v_0) = \frac{R_{yx,\text{raw}}^{\downarrow}(B, v_0) - R_{yx,\text{raw}}^{\uparrow}(-B, v_0)}{2}$$

$$R_{yx}^{\uparrow}(-B, v_0) = -R_{yx}^{\downarrow}(B, v_0), R_{yx}^{\downarrow}(-B, v_0) = -R_{yx}^{\uparrow}(B, v_0)$$

Fig. 1

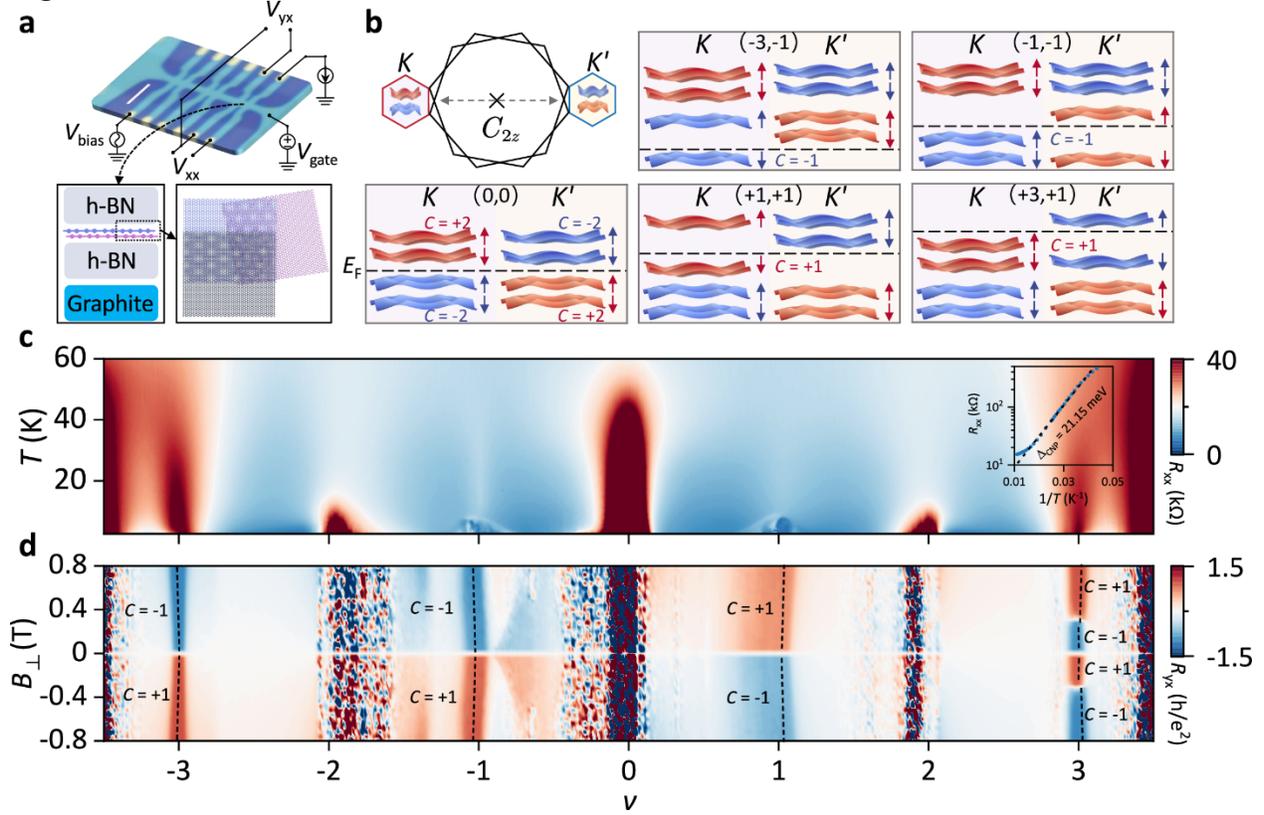

**Fig. 1 | Chern insulators in MATBG aligned to hBN. a,** Optical microscopy image (scale bar is 5μm) of device D1 and measurement configuration. Four terminal transport measurements for longitudinal resistance $R_{xx}$ and Hall resistance $R_{yx}$ are achieved by applying an alternating excitation $V_{bias}$ and measuring the voltage drop $V_{xx}$, $V_{yx}$ and the current $I$ flowing through the device (top panel). The cross-section and schematic display that the MATBG with hBN encapsulation is further aligned with top hBN and gated with graphite (bottom panel). **b,** Mini-Brillouin zone of MATBG with $C_{2z}$ symmetry breaking and different Chern band configurations with (ν, C) indices of (0, 0), (-3, -1), (-1, -1), (+1, +1) and (+3, +1). **c,** Temperature dependence of $R_{xx}$ (measured in a cryogenic refrigerator with a base temperature of 2.7 K) in device D1, the illustration in the upper right corner shows the Arrhenius plot for ν = 0 (CNP), with an extracted gap of $\triangle_{CNP}$ = 21.15 meV. The activation energy gap at the CNP is very large, strongly suggesting a perfect alignment between MATBG and the hBN substrate. **d,** Anti-symmetrized (see Methods for details) $R_{yx}$ versus filling factor ν and out-of-plane magnetic field $B_\perp$ obtained at T = 10 mK in device D1. The slanted dashed lines in the figure represent the evolution of the Chern insulating states (-3, -1), (-1, -1), (+1, +1) and (+3, +1) with the magnetic field according to the Streda formula.

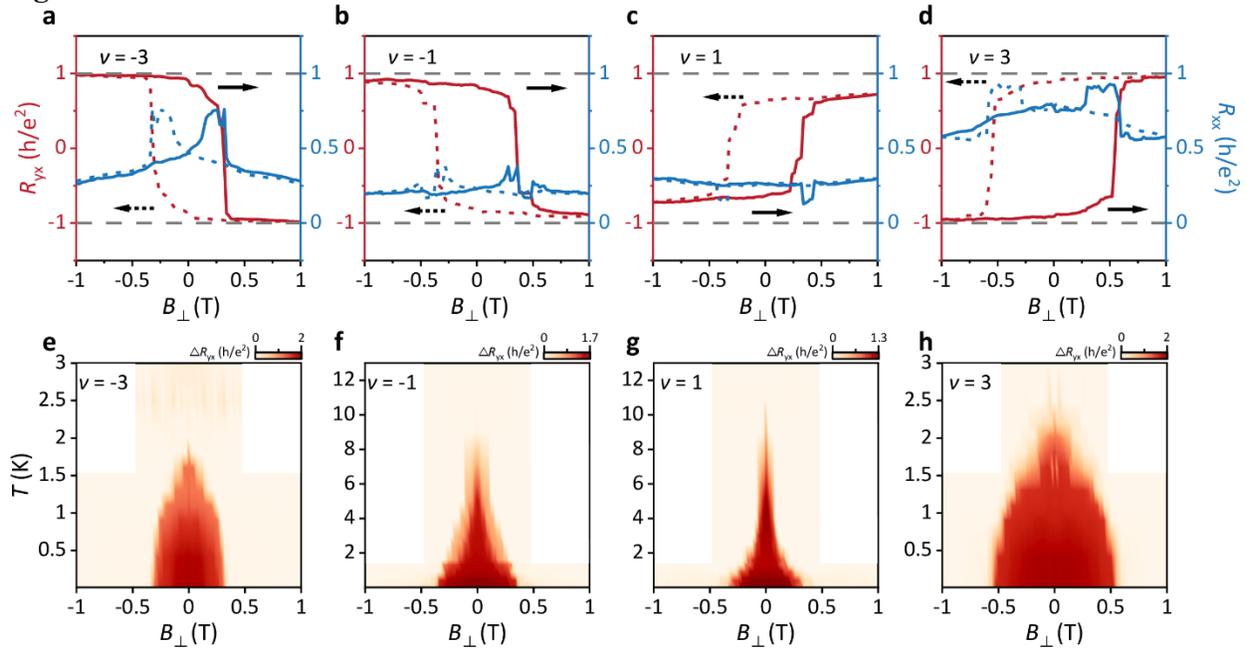

**Fig. 2 | Magnetic hysteresis loops. a-d,** Symmetrized $R_{xx}$ and anti-symmetrized $R_{yx}$ as a function of $B_\perp$ measured at $v = \pm 1$ and $\pm 3$ states, $T = 10$ mK. Dashed and solid lines correspond to sweeping the $B_\perp$ field back and forth indicated by the arrows. **e-h,** $\Delta R_{yx} = |R_{yx}(B_\perp$ sweeping up$) - R_{yx}(B_\perp$ sweeping down$)|$ versus $B_\perp$ field at different temperatures measured at $v = \pm 1$ and $\pm 3$ states.

Fig. 3

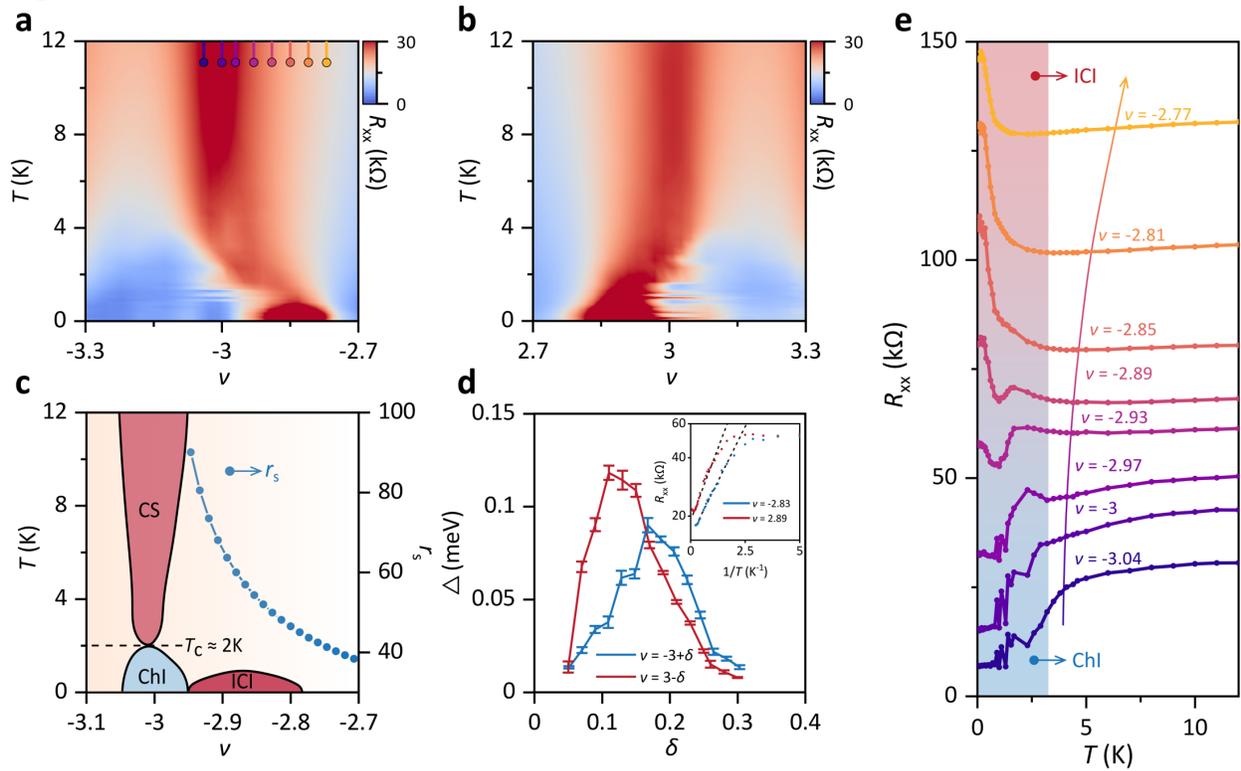

**Fig. 3 | Incommensurate correlated insulator (ICI) states near $v = \pm 3$. a,** and **b,** $R_{xx}$ color maps as a function of temperature $T$ and $v$ near the $v = -3$ and $v = +3$ states at zero magnetic field. **c,** Schematic diagram illustrating the different correlated phases near $v = -3$, extracted from **a** and the Wigner-Seitz radius $r_s$ obtained from theoretical calculations, as a function of the filling factor $v$ (right y-axis). **d,** Activation energy gap $\triangle_{gap}$ as a function of the offset $\delta$ of the filling factor $v$, where $v$ is near $3-\delta$ (shown in red) and $-3+\delta$ (shown in blue) corresponds to insulating behavior. The illustration in the upper right corner shows the Arrhenius plot for $v = -2.83$ and $2.89$. **e,** Temperature-dependent resistance $R_{xx}$ near $v = -3$ state for a series of selected moiré fillings indicated by the short lines in **a** (the line cut curves are offset for better clarification). The blue region represents the Chern insulators showing metallic behavior in $R_{xx}$ nevertheless due to the existence of edge states while the red region represents the ICI states showing insulating behavior in $R_{xx}$.

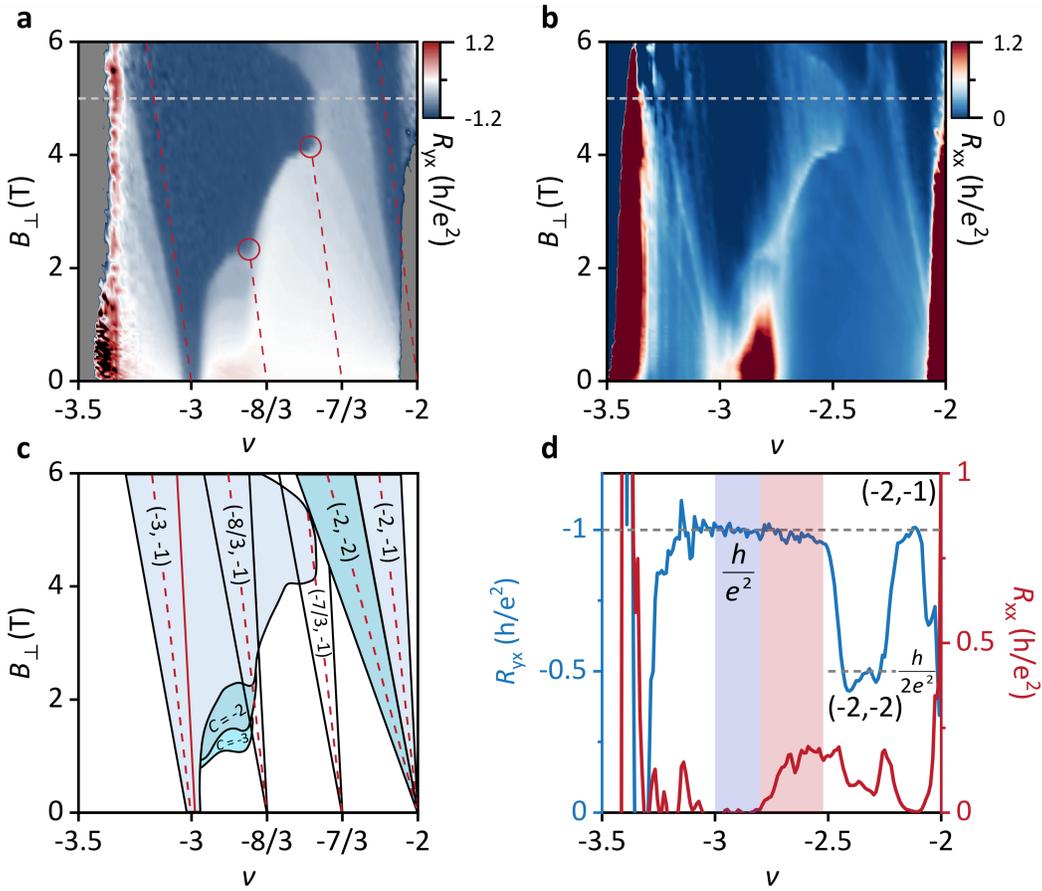

**Fig. 4 | Topological incommensurate Chern insulator (IChI) states at $v = -3+\delta$. a-b,** Unsymmetrized Landau fan diagram of the $R_{yx}$ (Anti-symmetrized $R_{yx}$ is shown in Extended Fig. 8) and $R_{xx}$. The red dashed lines represent the evolution of Chern insulator states with Chern number $C = -1$ emanating from $v = -3, -8/3, -7/3$, and $-2$ based on the Streda formula at $T = 10$ mK. **c,** Schematic diagram indicating the Chern insulator states with $(v, C)$ indices of $(-3, -1)$, $(-8/3, -1)$, $(-7/3, -1)$, $(-2, -2)$ and $(-2, -1)$ extracted from the Landau fan diagram **a**. **d,** Line cuts of $R_{xx}$ and $R_{yx}$ at $B_\perp = 5$ T (corresponding to the gray dashed lines in **a** and **b**). The purple region indicates where $R_{xx}$ vanishes completely to zero, while the pink region signifies where $R_{xx}$ begins to exhibit non-zero values. In both regions, $R_{yx}$ shows quantized values of $-h/e^2$.

Fig. 5

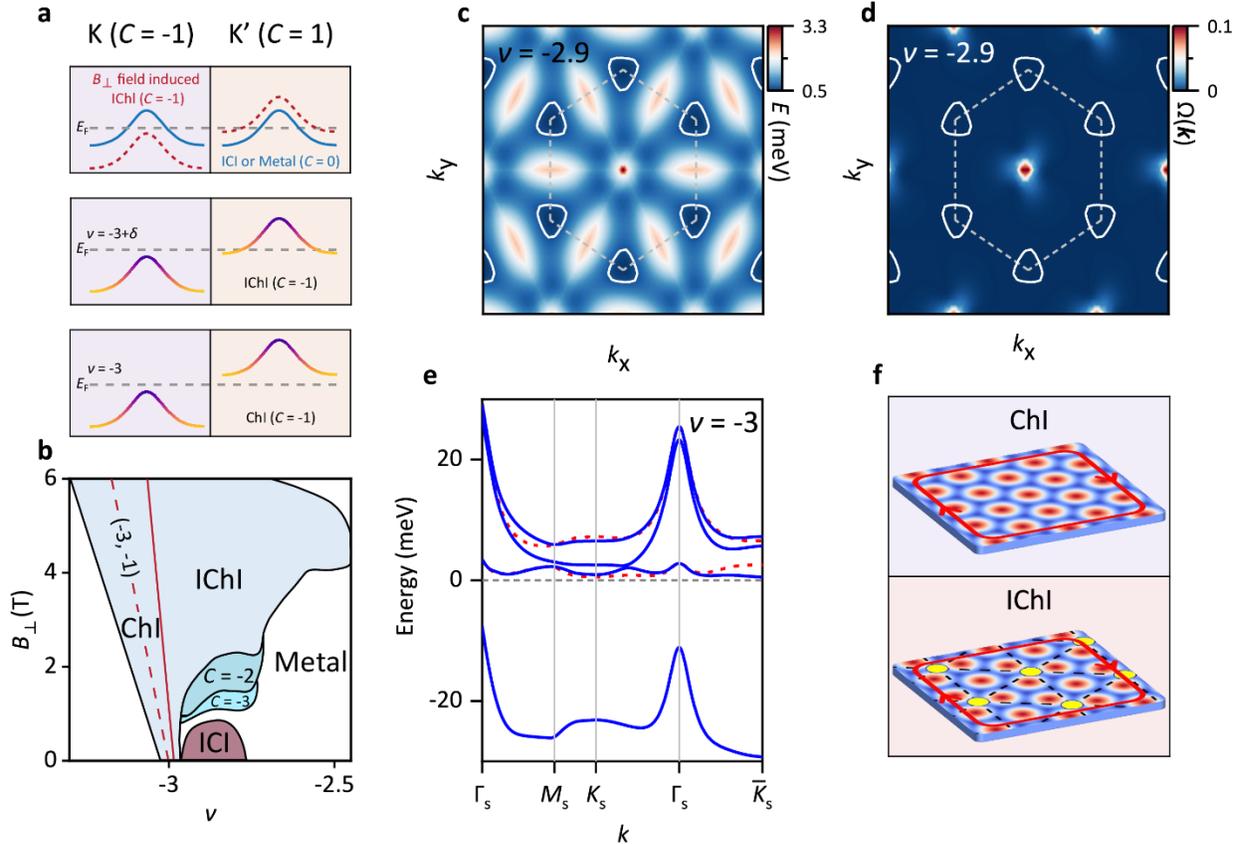

**Fig. 5 | Theoretical modeling and numerical calculation for topological IChI states. a,** Schematic illustration of ground states of ChI, ICI (or metal) and IChI, the colors in the middle and bottom figures represent the distribution of the Berry curvature qualitatively, the Berry curvature gradually increases from yellow to purple. **b,** Schematic phase diagram illustrates different correlated phases near $v = -3$. It is noteworthy that the region representing IChI states can extend to zero magnetic field (zero-filed IChI state). As the magnetic field increases, this region progressively expands; however, above ~ 5 T, its extent gradually diminishes. **c,** Band structure of the unoccupied flat band above the $v = -3$ gap and Fermi surface when the filling factor is $v = -2.9$. The gray dashed lines show the moiré Brillouin zone and the white lines represent the Fermi surface. **d,** Distribution of the Berry curvature in the unoccupied flat band above the $v = -3$ gap. The Chern number of the unoccupied band is $C = +1$. The hot spots of the Berry curvature are localized near the $\Gamma_s$ point, while becoming very small near $K_s$ and $M_s$ points. The gray dashed lines also show the moiré Brillouin zone, while the white lines represent the same Fermi surface as shown in **c**. **e,** Band structure of the magic-angle TBG at the filling factor of $v = -3$. The red dashed lines and the blue lines denote the energy bands from two atomic valleys. **f,** Real-space illustration of the ChI and IChI states. In the upper diagram, each moiré supercell should be filled with three holes, corresponding to $v = -3$. In the bottom diagram, the filling factor slightly deviates from the integer filling, i.e., $v = -3+\delta$, where the yellow dots represent the additional electrons relative to the $v = -3$ state.

**Extended Data Fig. 1**

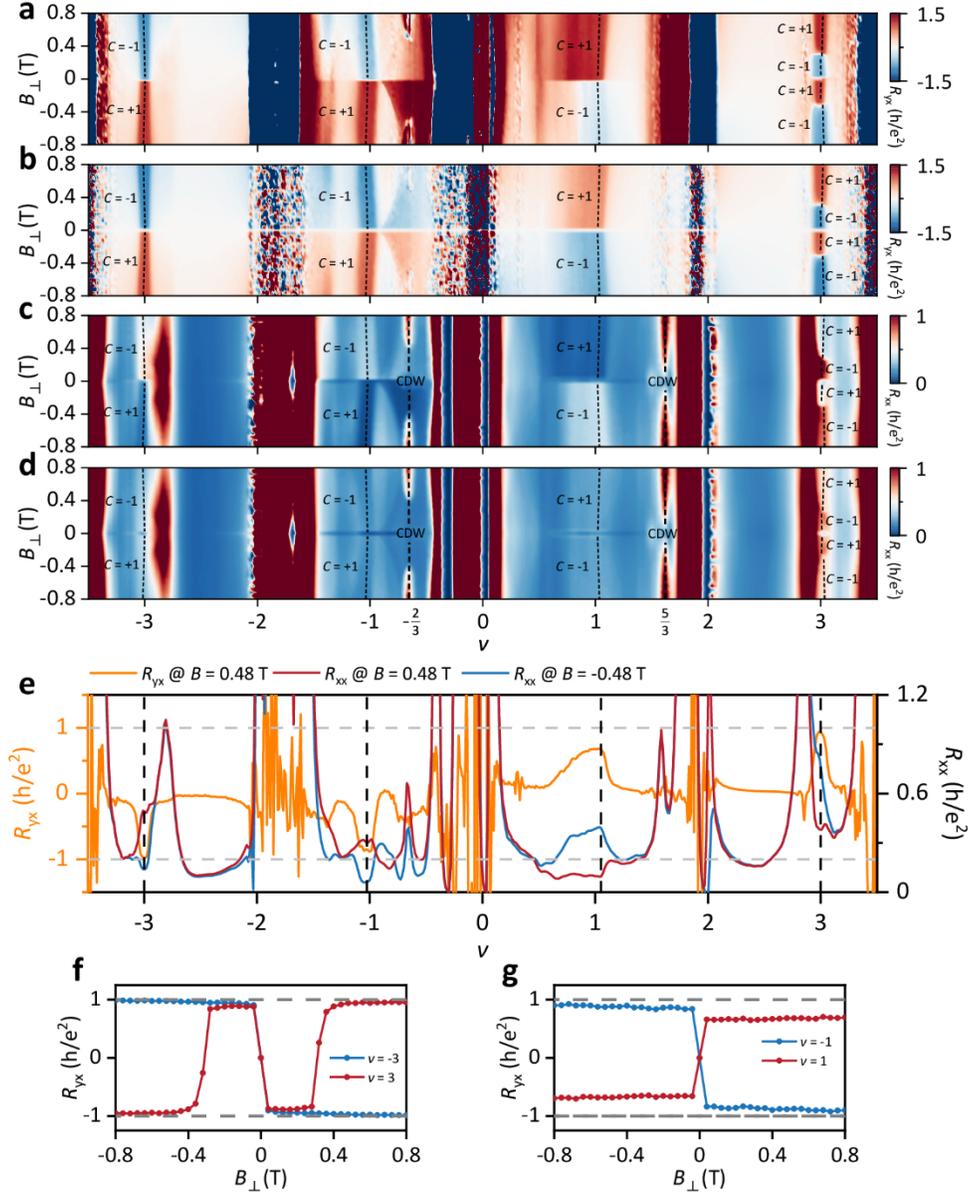

**Extended Data Fig. 1 | Landau fan diagrams at low $B_\perp$ magnetic fields. a** and **b,** Unsymmetrized (**a**) and anti-symmetrized (**b**) $R_{yx}$ versus filling factor $v$ and out-of-plane magnetic field $B_\perp$ obtained at $T = 10$ mK. **c** and **d,** Unsymmetrized (**c**) and symmetrized (**d**) $R_{xx}$ versus $v$ and $B_\perp$. The slanted dashed lines in the figure represent the evolution of the Chern insulating states with the magnetic field according to the Streda formula. Additionally, as the magnetic field increases, $R_{xx}$ shows pronounced resistance peaks at filling factors $v = -2/3$ and $v = 5/3$, indicating the formation of emerging CDW states, as represented by vertical dashed lines in **c** and **d**. **e,** Line cuts of anti-symmetrized $R_{yx}$ and unsymmetrized $R_{xx}$ versus $v$ at $B_\perp = \pm 0.48$ T. $R_{xx}$ exhibits distinct dips at $v = 1$ and $v = 3$ when subjected to $B_\perp = +0.48$ T, and similar dips are observed at $v = -1$ and $v = -3$ under $B_\perp = -0.48$ T. Meanwhile, $R_{yx}$ presents nearly quantized values at $v = \pm 3$ and $v = \pm 1$. **f** and **g,** Anti-symmetrized $R_{yx}$ versus out-of-plane magnetic field $B_\perp$ at $v = \pm 3$ (**f**) and $v = \pm 1$ (**g**) as extracted from Fig. **1d**.

**Extended Data Fig. 2**

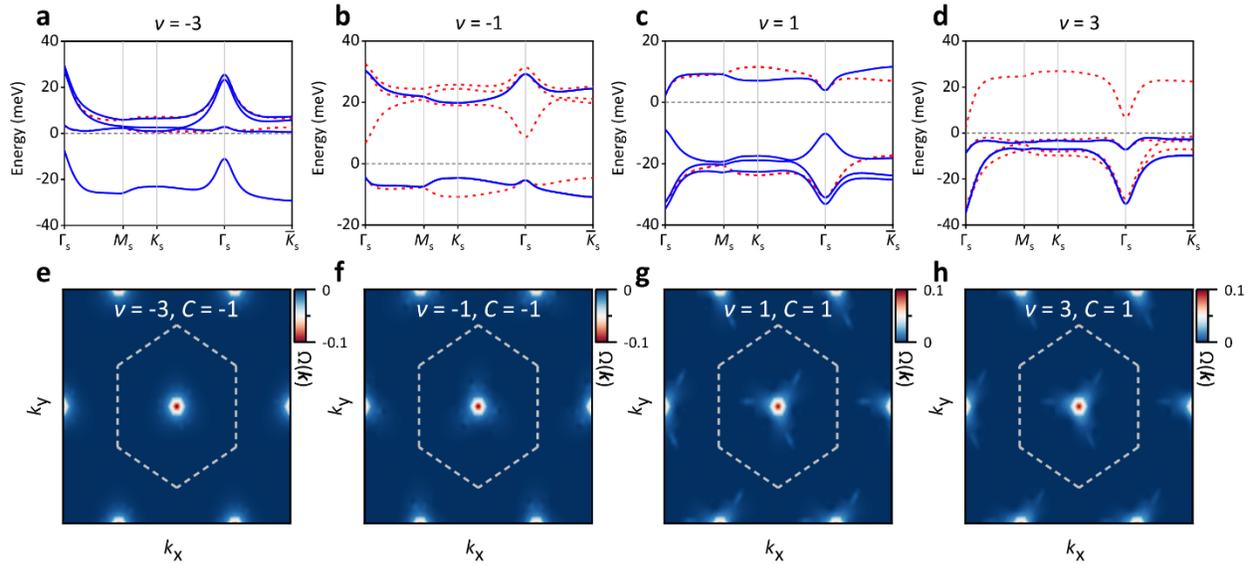

**Extended Data Fig. 2 | Energy band, Berry curvature and Chern number at the $v = \pm 3$ and $v = \pm 1$ states obtained from theoretical calculations. a-d**, Band structure of the magic-angle TBG at the filling factor of $v = -3, -1, 1$ and $3$. The red dashed lines and the blue lines denote the energy bands from two atomic valleys. **e-h,** Distribution of the Berry curvature in the occupied flat band below the $v = -3, -1, 1$ and $3$ gap. The gray dashed lines show the moiré Brillouin zone. The Chern number of the occupied bands is $C = +1$ for $v = +3$ and $v = +1$ under zero magnetic field and is $C = -1$ for $v = -3$ and $v = -1$.

**Extended Data Fig. 3**

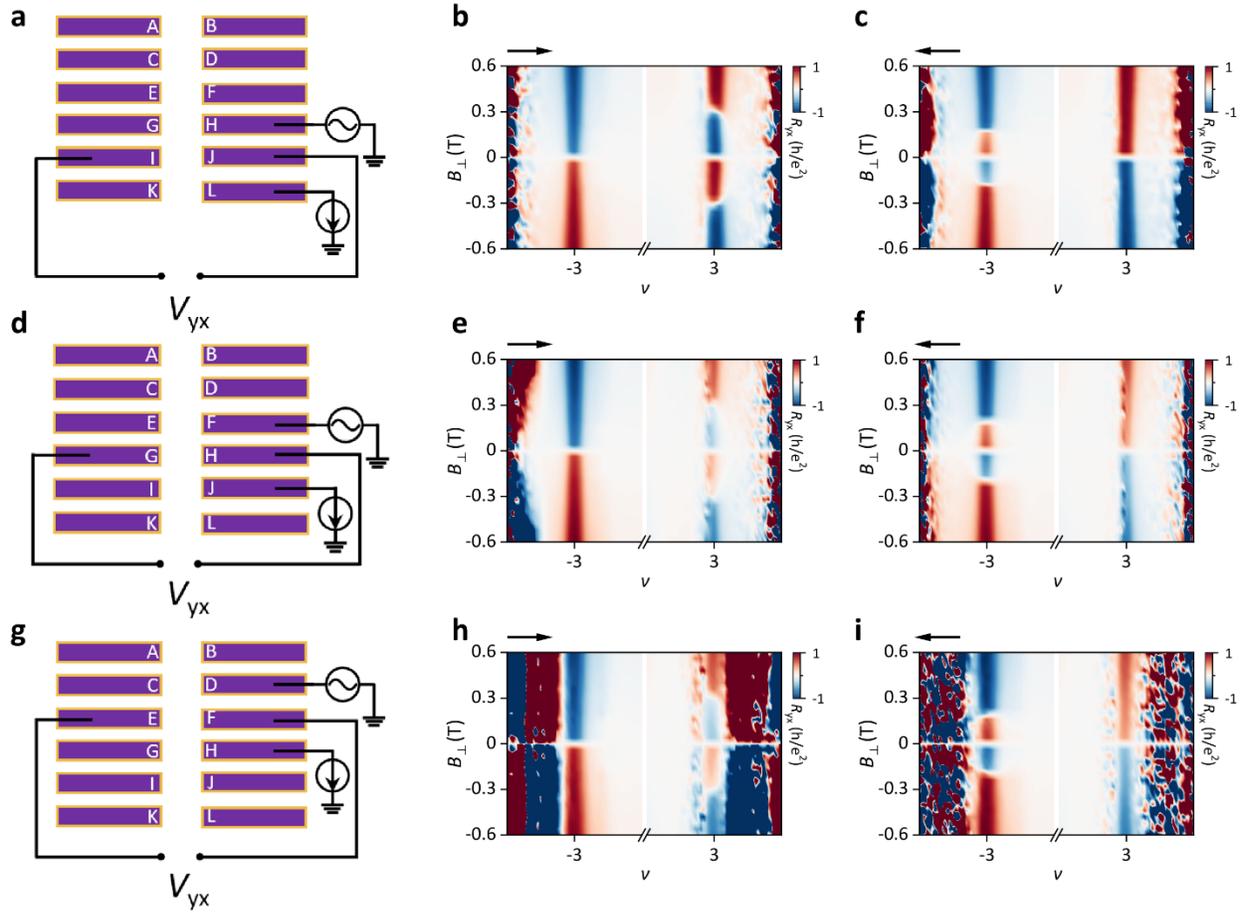

**Extended Data Fig. 3 | Magnetic states determined by the gate sweeping direction. a, d,** and **g,** Measurement configurations for **b-c, e-f,** and **h-i** respectively. **b, e, and h,** Anti-symmetrized $R_{yx}$ versus filling factor $v$ and out-of-plane magnetic field $B_\perp$ obtained at $T = 10$ mK when scanning the back gate from negative to positive voltage. **c, f, and i,** Anti-symmetrized $R_{yx}$ versus $v$ and $B_\perp$ when scanning the back gate from positive to negative voltage. The arrows in the top left corner of the figure indicate the direction of the back gate scan. Upon scanning the back gate voltage from negative to positive, an additional sign reversal of $R_{yx}$ is observed at the $v = 3$ state in the presence of a few hundred milli-Tesla magnetic field $B_\perp$. Conversely, when scanning from positive to negative voltage, the sign reversal occurs at the $v = -3$ state.

**Extended Data Fig. 4**

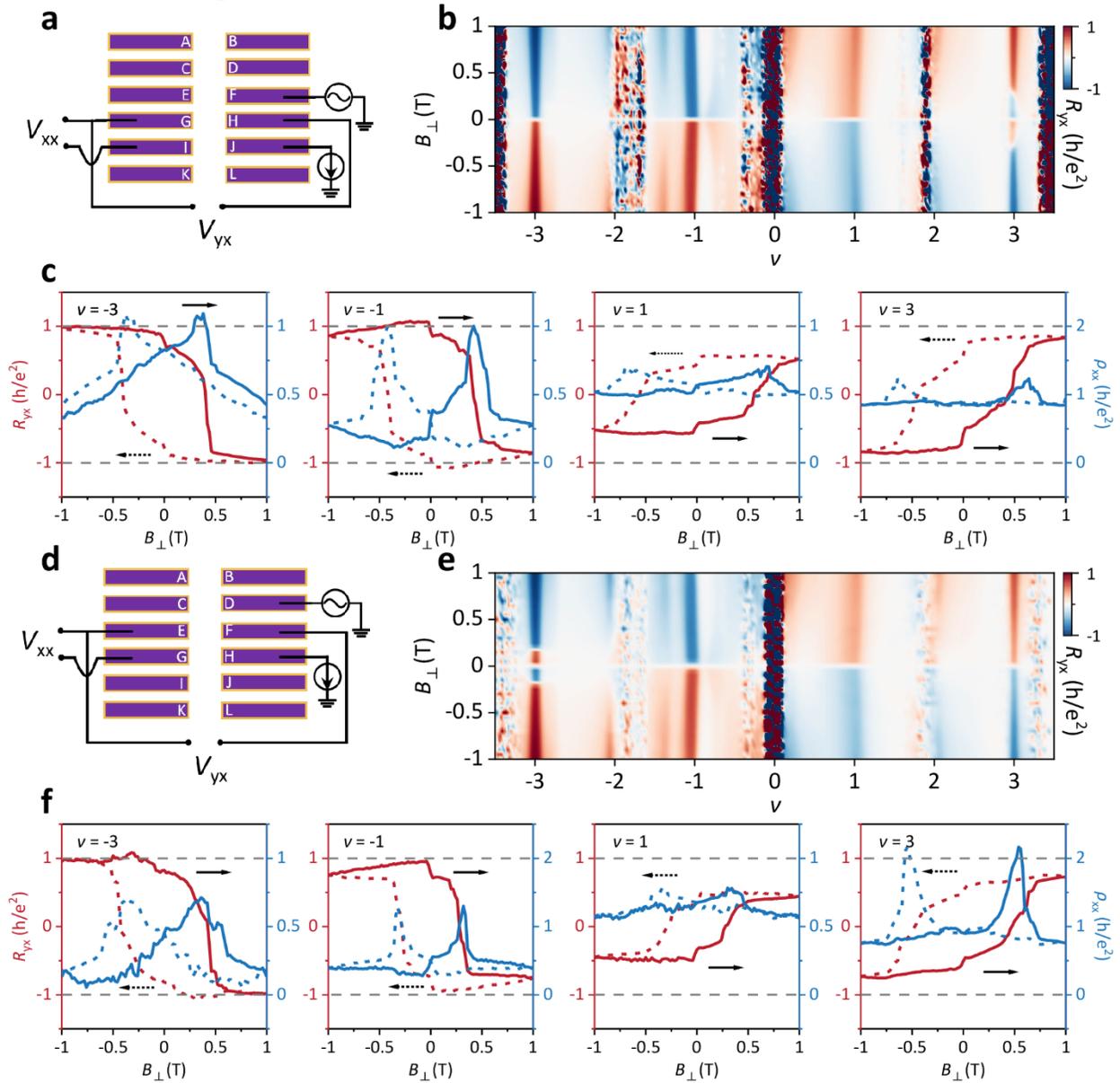

**Extended Data Fig. 4 | Hysteresis loops and Landau fan diagrams of other Hall bar contact pairs in device D1. a** and **d,** Measurement configurations for **b-c** and **e-f** respectively. **b,** and **e,** Anti-symmetrized $R_{yx}$ versus $v$ and $B_\perp$. **c,** and **f,** Symmetrized longitudinal resistivity $\rho_{xx}$ and anti-symmetrized $R_{yx}$ as a function of $B_\perp$ measured at $v = \pm 1$ and $\pm 3$ states. Dashed and solid lines correspond to sweeping the $B_\perp$ field back and forth indicated by the arrows. The all data are acquired at $T = 10$ mK.

**Extended Data Fig. 5**

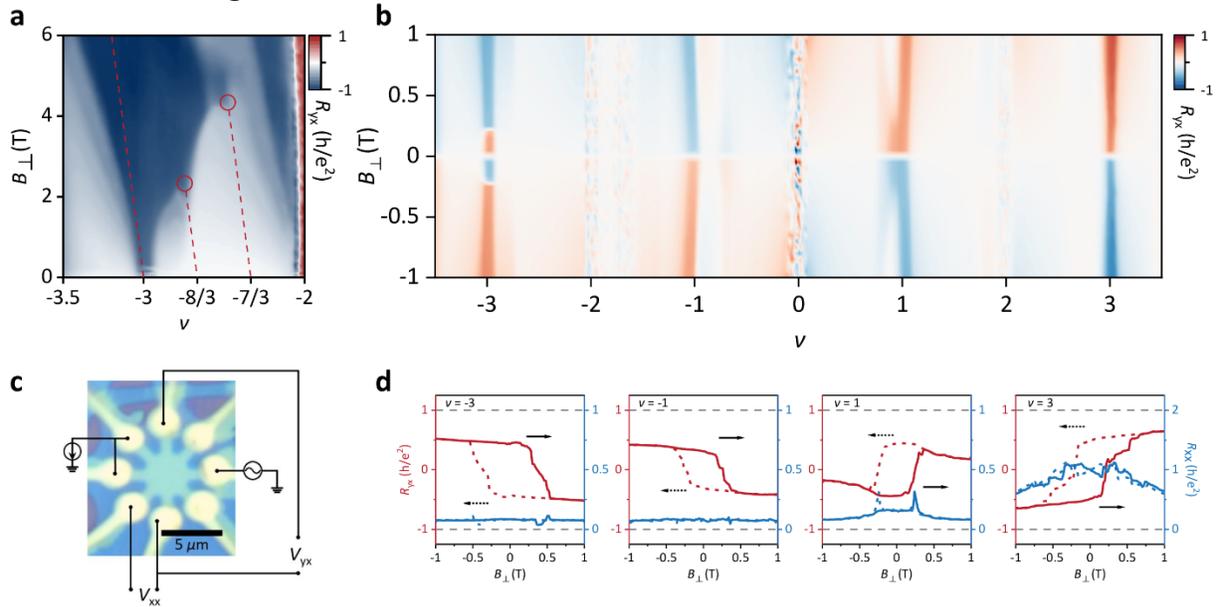

**Extended Data Fig. 5 | Hysteresis loops and Landau fan diagrams of MATBG/h-BN device D2. a,** Unsymmetrized $R_{yx}$ versus $v$ and $B_\perp$ (0 ~ 6 T). **b,** Anti-symmetrized $R_{yx}$ versus $v$ and $B_\perp$ (-1 ~ 1 T). **c,** Optical microscopy image (scale bar is 5 $\mu$m) of device D2 and measurement configuration. **d,** Symmetrized longitudinal resistance $R_{xx}$ and anti-symmetrized Hall resistance $R_{yx}$ as a function of $B_\perp$ measured at $v = \pm1$ and $\pm3$ states. Dashed and solid lines correspond to sweeping the $B_\perp$ field back and forth indicated by the arrows. The all data are acquired at $T = 10$ mK. Which indicates the reproducibility of the commensurate and incommensurate Chern insulator states across different devices.

**Extended Data Fig. 6**

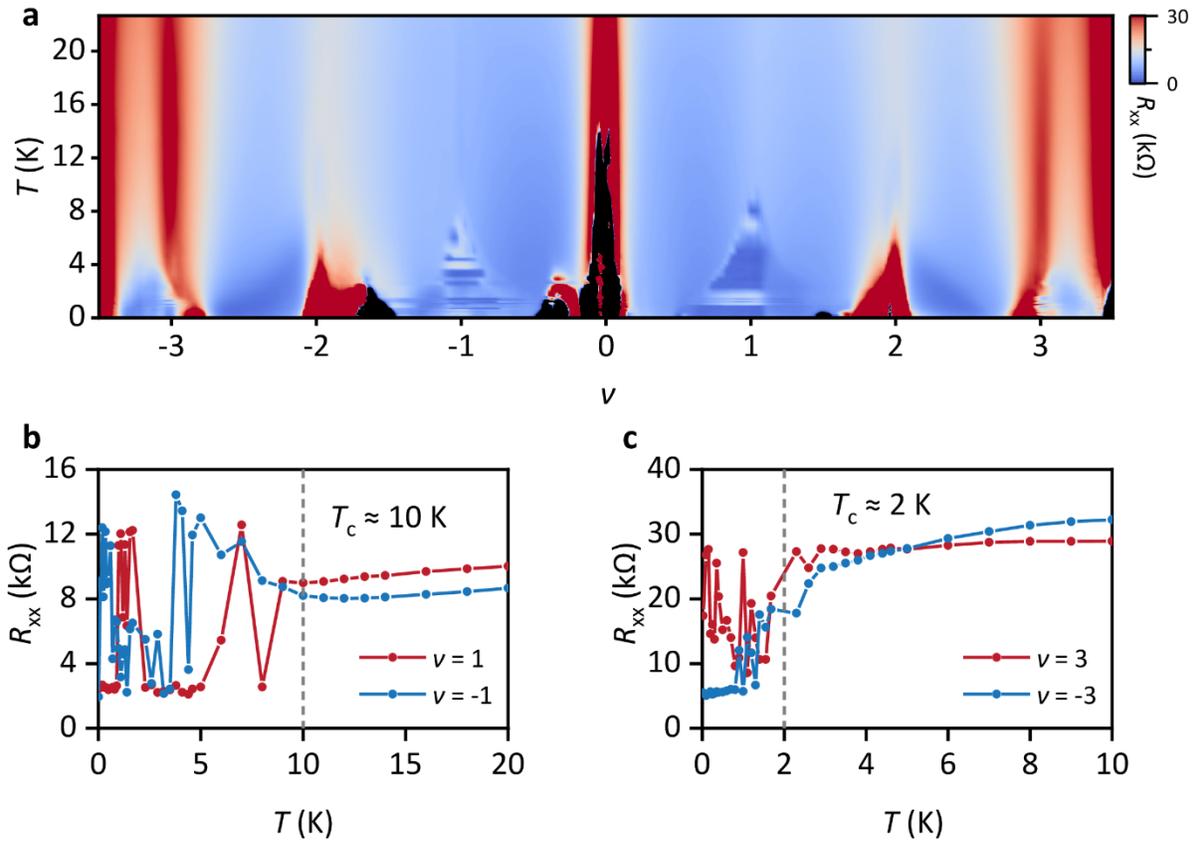

**Extended Data Fig. 6 | Longitudinal resistance at different temperatures and filing factors.**
**a,** Colormap $R_{xx}$ (measured at zero magnetic field) against $v$ and $T$. **b,** Linecut traces of $R_{xx}$ versus $T$ at $v = \pm 1$ from **a**. **c,** Linecut traces of $R_{xx}$ versus $T$ at $v = \pm 3$ from **a.**

**Extended Data Fig. 7**

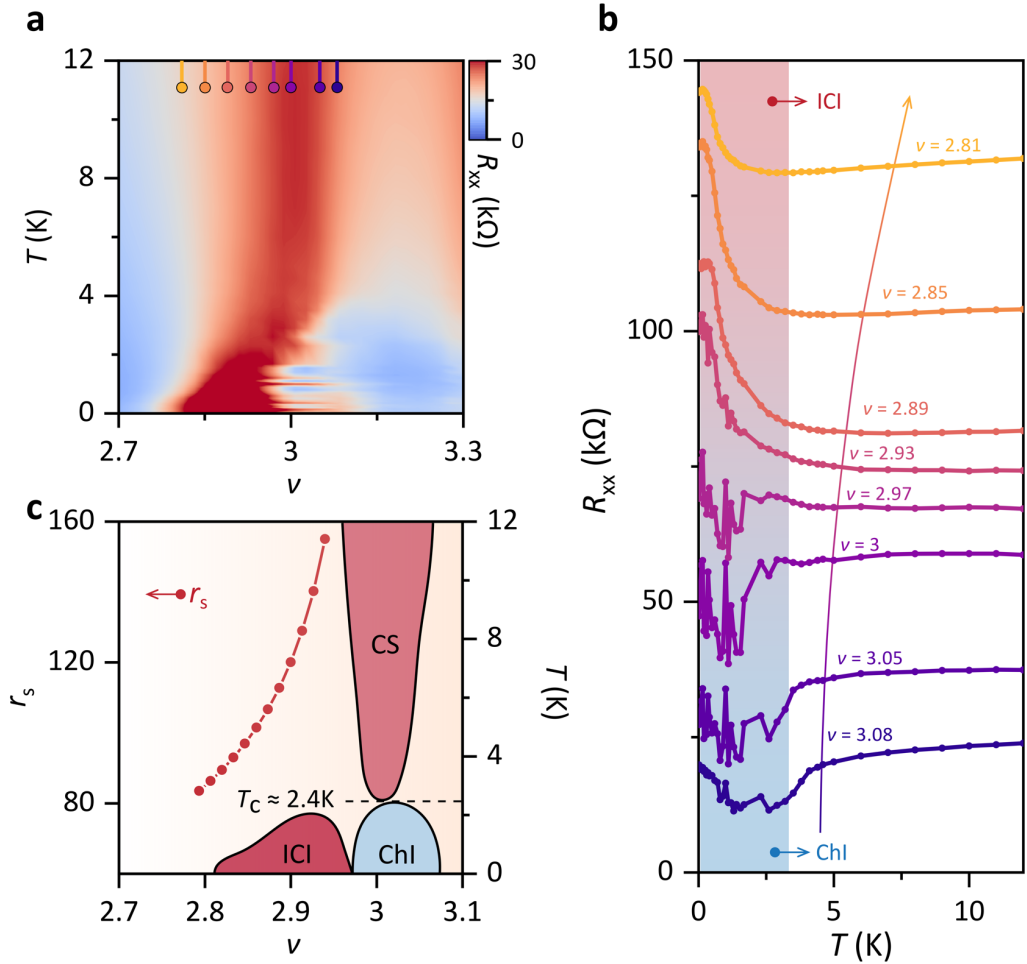

**Extended Data Fig. 7 | Incommensurate correlated insulator (ICI) states at $v = 3-\delta$. a,** $R_{xx}$ color maps as a function of temperature $T$ and $v$ near the $v = +3$ states at zero magnetic field. **b,** Schematic diagram illustrating the different correlated phases near $v = +3$, extracted from **a** and the Wigner-Seitz radius $r_s$ obtained from theoretical calculations, as a function of the filling factor $v$ (left y-axis). **c,** Temperature-dependent resistance $R_{xx}$ near $v = +3$ state for a series of selected moiré fillings indicated by the short lines in **a** (the line cut curves are offset for better clarification). The blue region represents the Chern insulators showing metallic behavior in $R_{xx}$ nevertheless due to the existence of edge states while the red region represents the ICI states showing insulating behavior in $R_{xx}$.

**Extended Data Fig. 8**

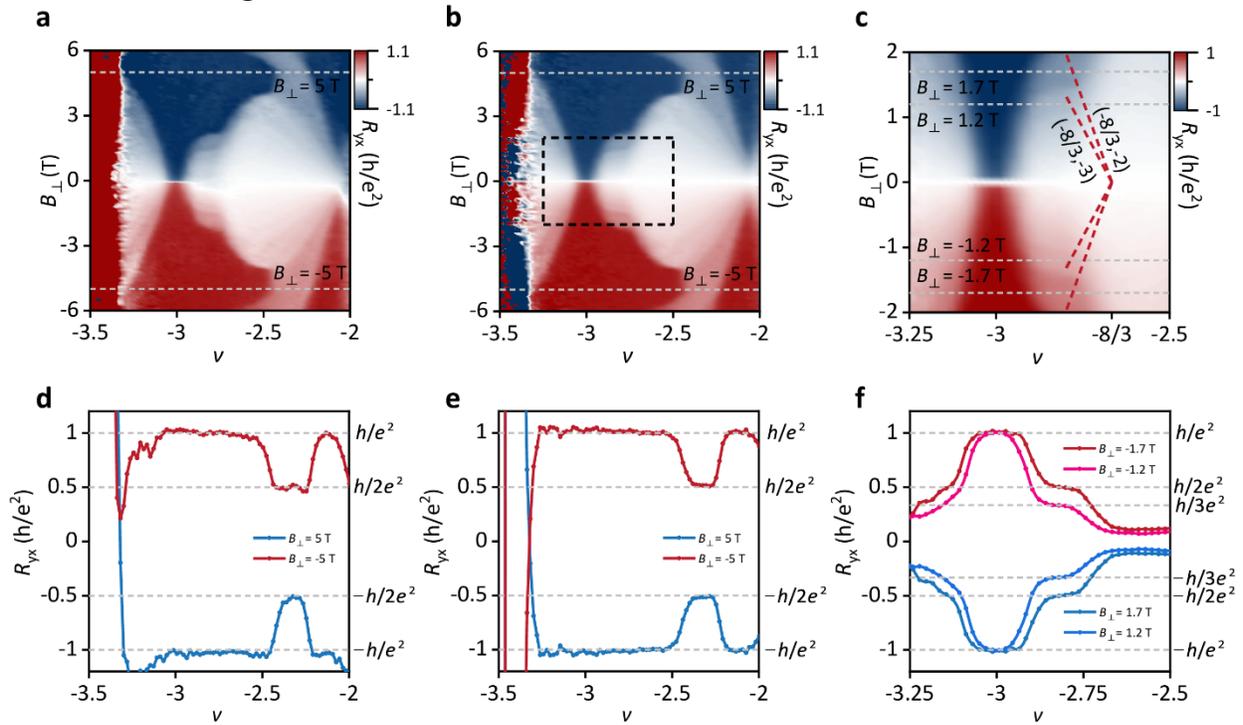

**Extended Data Fig. 8 | Landau fan diagrams near $v = -3$. a** and **b,** Unsymmetrized (**a**) and anti-symmetrized (**b**) $R_{yx}$ versus filling factor $v$ and out-of-plane magnetic field $B_\perp$ obtained at $T = 10$ mK. **c,** Anti-symmetrized $R_{yx}$ map obtained in the boxed region in **b**. The slanted dashed lines in the figure represent the evolution of the Chern insulating states (-8/3, -2) and (-8/3, -3) with the magnetic field according to the Streda formula. **d-f,** Line cuts of $R_{yx}$ corresponding to the gray dashed lines in **a-c** respectively.